\documentclass[%
 reprint,
superscriptaddress,
 amsmath,amssymb,
 aps,
]{revtex4-2}

\usepackage{graphicx}% Include figure files
\usepackage{dcolumn}% Align table columns on decimal point
\usepackage{bm}% bold math
\usepackage{chemformula} % Formula subscripts using \ch{}
\usepackage[T1]{fontenc} % Use modern font encodings
\usepackage{siunitx}
\usepackage{comment}
\usepackage{graphicx}% Include figure files
\usepackage{dcolumn}% Align table columns on decimal point
\usepackage{bm}% bold math
\usepackage{empheq}
\usepackage{siunitx}
\DeclareSIUnit{\dBm}{dBm}
\usepackage{xcolor}

\usepackage{fixltx2e}
\usepackage{soul}
\usepackage{float}
\usepackage[normalem]{ulem}
\begin{document}

\preprint{APS/123-QED}

\title{Exploring electrochemical methods for 2D precision stress control in nanoscale devices}

\author{Di Chen}
\email{di.chen@bristol.ac.uk}
\affiliation{School of Physics, H.H. Wills Physics Laboratory, University of Bristol, Bristol BS8 1TL, United Kingdom}

\author{Natasa Vasiljevic}
\affiliation{School of Physics, H.H. Wills Physics Laboratory, University of Bristol, Bristol BS8 1TL, United Kingdom}
\author{Andrei Sarua}
\affiliation{Centre for Device Reliability and Thermography, School of Physics, H.H. Wills Physics Laboratory, University of Bristol, Bristol BS8 1TL, United Kingdom}
\author{Martin Kuball}
\affiliation{Centre for Device Reliability and Thermography, School of Physics, H.H. Wills Physics Laboratory, University of Bristol, Bristol BS8 1TL, United Kingdom}

\author{Krishna C. Balram}
\email{krishna.coimbatorebalram@bristol.ac.uk}
\affiliation{Quantum Engineering Technology Labs and School of Electrical, Electronic and Mechanical Engineering, University of Bristol,
Bristol BS8 1UB, United Kingdom}

\date{\today}% It is always \today, today,

\begin{abstract}
Tuning the local film stress (and associated strain) provides a universal route towards exerting dynamic control on propagating fields in nanoscale geometries, and engineering controlled interactions between them. The majority of existing techniques are adapted for engineering either uniform stresses or fixed stress gradients, but there is a need to develop methods that can provide independent precision control over the local stress at the nanoscale in the 2D plane. Here, we explore electrochemical absorption of hydrogen in structured palladium thin-film electrodes, and the associated shape-dependent stress to engineer controlled, localized stresses in thin films. We discuss the prospects of this technique for precision dynamic tuning of nanoscale opto-electro-mechanical devices and the development of field-programmable non-volatile \textit{set-and-forget} architectures. We also outline some of the key challenges that need to be addressed with a view towards incorporating electrochemical stress tuning methods for post-processing foundry devices.   
\end{abstract}

\maketitle

\section{Introduction}

Exerting exquisite control over propagating fields in nanoscale (sub-\qty{}{\um}) geometries using stress (and the associated strain) has been a unifying theme across diverse research fields, ranging from microelectronics \cite{chu2009strain} and integrated photonics \cite{tian2024piezoelectric} to microwave acoustics \cite{c2024piezoelectric}. The two hallmark examples of its impact on modern life are the use of stress to improve threshold current density in semiconductor quantum well lasers that power modern fiber-optic communication \cite{adams2011strained}, and improving channel carrier mobility in nanoscale transistors \cite{thompson200490}. These examples also illustrate how stress provides a universal tuning knob that can simultaneously manipulate multiple fields at the nanoscale, with a view towards controlling interactions in multi-resonant geometries, as is the case in devices such as piezoelectric microwave to optical quantum transducers \cite{balram2022piezoelectric}, that rely on resonant acousto-optic interactions in nanoscale cavities. 

\begin{figure}[htbp]
    \centering
    \includegraphics[width=1\linewidth]{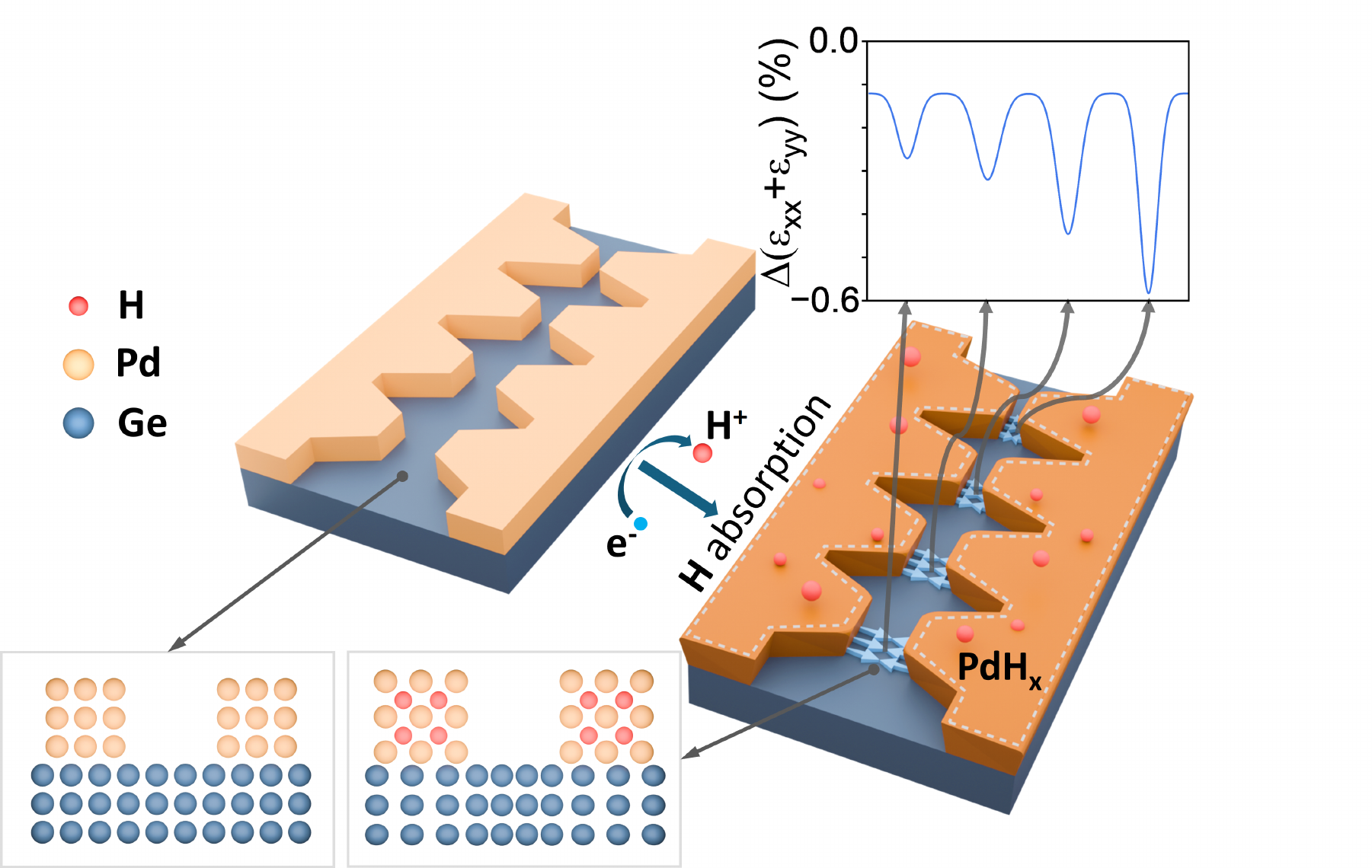}
    \caption{Schematic illustration of electrochemical absorption induced localized stress generation. Pd electrodes are lithographically patterned on a Ge-on-Si substrate. Hydrogen is electrochemically absorbed into the Pd electrodes, resulting in a hydrostatic volume expansion which scales with the [H/Pd] ratio. This expansion exerts a localized stress on the underlying Ge film and its magnitude can be controlled by choosing the electrode shape. The inset (top-right) illustrates how the compressive stress registered at the centre between the electrodes scales with reducing electrode gap.} 
    \label{Figure 1}
\end{figure}

\begin{figure*}[thbp]
    \centering
    \includegraphics[width=1\linewidth]{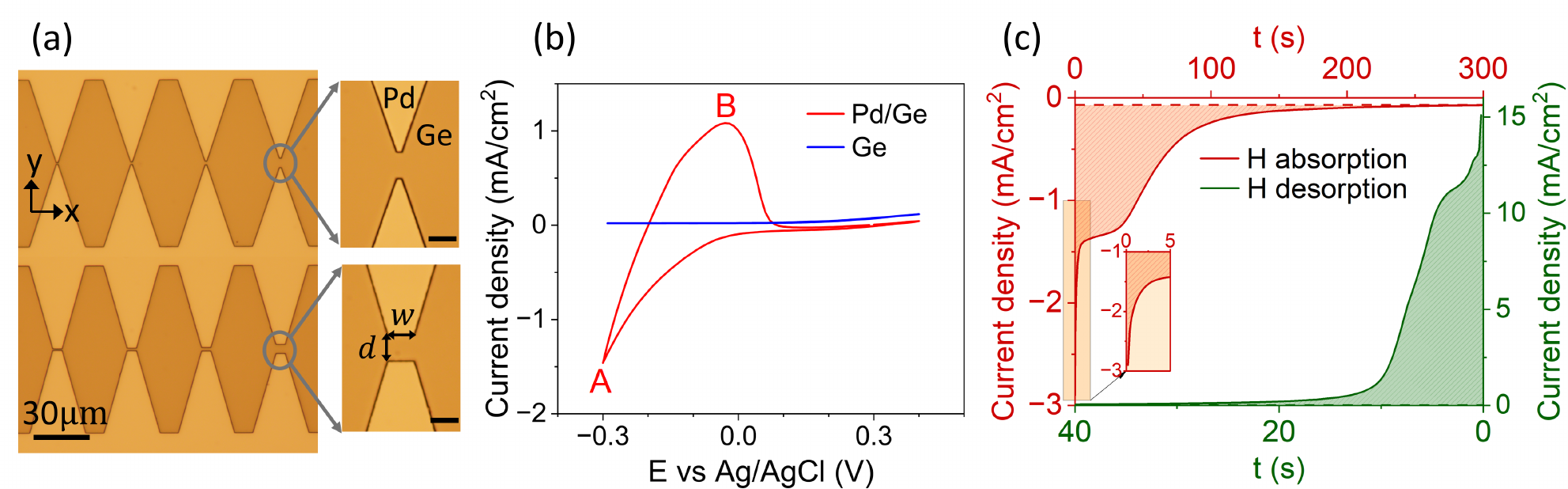}
    \caption{
    (a)  Optical microscope images of structured \qty{200}{\nm}-thick Pd electrodes on Ge-on-Si substrate (scale bar for zoomed-in insets is \qty{5}{\um}); (b) Cyclic voltammetry (CV) curves for Ge-on-Si with and without Pd electrode in \qty{0.1}{M} \ch{H2SO4} (scan rate of \qty{50}{\milli\volt\per\second}); (c) Chronoamperometry (CA) curves for H-loading at \qty{-0.26}{\volt} (red, left y-axis and top x-axis) and desorption (green, right y-axis and bottom x-axis) at \qty{0.24}{\volt} with dashed red and green lines indicating the background currents.} 
    \label{Figure 2}
\end{figure*}

A common feature amongst all existing approaches to applying stress in integrated platforms, whether actuated using thermal, piezoelectric \cite{wen2023high}, electrostatic \cite{ayan2019strain} or magnetostatic effects \cite{ekinci2005nanoelectromechanical}, is that they are well adapted to applying uniform stresses or fixed stress gradients. But there are scenarios in which one ideally wants to control the 2D stress pattern (and the associated strain) at the nanoscale. These include the use of nansocale strain to create giant pseudo-magnetic fields in graphene \cite{levy2010strain} and overcoming nano-fabrication induced variations \cite{wong2004strain, hatipoglu2024situ, grim2019scalable} in coupled cavity arrays for exploring many-body physics analogs of condensed matter phenomena \cite{hartmann2008quantum}. The ultimate goal is to be able to precisely control both the magnitude and the sign (whether compressive or tensile) of the local stress at a given location with the flexibility to extend this spatial control arbitrarily over the 2D plane. The limitations of existing approaches mainly stem from either residual crosstalk (thermal effects) or the footprint needed for the actuator (almost all of the nanoelectromechanical systems approaches). A second related theme is the need for methods that can implement \textit{set-and-forget} tuning approaches that alleviate the need for an active tuning field. This becomes especially important when one is tuning large-scale resonant systems, a recurring theme in experimental implementations of quantum computing at scale in power or footprint constrained cryogenic environments.

Here, we explore the use of localized electrochemical absorption of hydrogen (H) in structured palladium (Pd) electrodes as a way to achieve localized control of stress with flexibility over engineering the stress pattern over the 2D plane. This enables the prospect of engineering field programmable \textit{set-and-forget} architectures with a view towards large scale control of resonant nano opto-electro-mechanical (NOEM) systems without requiring active tuning of each element. While the electrochemical absorption of hydrogen in palladium films is well understood \cite{jewell2006review}, this work builds on two key observations: (a) the hydrogen loading fraction in the electrode can in principle be controlled with high precision and as that determines the hydrostatic volume expansion directly, one has a way to program local strains precisely, (b) the spatial profile can be controlled by shaping the electrode through standard lithographic techniques. From a footprint perspective, the shaped Pd electrode functions as a nanoscale stress actuator, and therefore provides a natural route to efficiently interface with NOEM devices. 

We use a thin film of germanium (Ge) (\qty{1.6}{\um}), epitaxially grown \cite{jain2011tensile} on a silicon (Si) substrate as the test platform for proof-of-principle experiments for validating our ideas. The choice of Ge is mainly motivated by the prospect of using large tensile strains ($\approx2\%$) to engineer a transition from being an indirect bandgap semiconductor to one with a direct bandgap in a silicon-compatible material platform \cite{jain2011tensile, nam2011strained} which has potential benefits from an optoelectronics perspective, but the idea can be applied to any material platform of interest. In a way, one of the main advantages of this stress tuning method is the generality with which it can be applied to any potential NOEM platform being studied. 

\section{Engineering stress through structured volume expansion}

Fig.\ref{Figure 1} shows a schematic illustration of the technique. Structured Pd electrodes ($t_{Pd}=$ \qty{200}{\nm}) are lithographically patterned using standard electron beam lithography and lift-off procedures. A representative set of fabricated devices are shown in Fig.\ref{Figure 2}(a).  We use a \qty{2.5}{\nm} chromium  (Cr) layer underneath to maintain adhesion of the Pd film to the underlying substrate. Hydrogen is then electrochemically absorbed into the electrodes resulting in hydrostatic volume expansion that can be controlled by the [H/Pd] ratio. Structuring the electrode shape allows us to exert local control over the stress, as shown in the inset of Fig.\ref{Figure 1} which shows how the stress at the centre of the electrode gap scales with reducing gap width. These two parameters, [H/Pd] ratio and the geometric shape provide two (independent) degrees of freedom for engineering the stress over the 2D plane. In the experiments reported here, the [H/Pd] loading is constant for all the shaped electrodes at any given instant, but one can easily extend this scheme to exert local control over the [H/Pd] ratio for each electrode by applying a different potential on each. 

Fig.\ref{Figure 2}(b) shows the cyclic voltammetry (CV) curves of Ge, with and without the Pd electrode, obtained in \qty{0.1}{M} \ch{H2SO4} at a scan rate of \qty{50}{\milli\volt\per\second}, using Ag/AgCl as the reference electrode and a Pt wire as the counter electrode. No peaks were observed for the bare Ge sample, indicating that no redox reaction occurred on the Ge surface. Peak A on the curve of the sample with the Pd electrode corresponds to hydrogen absorption (\ch{H^+ + e^- -> H_{abs}}) and hydrogen evolution reaction (\ch{2 H^+ + 2 e^- -> H2}), while peak B corresponds to the desorption (\ch{H_{abs} -> H^+ + e^-}) \cite{gabrielli2004investigation}. Before the stress induced due to H-loading was quantified, the sample was cycled until repeatable CV behaviour was obtained. Micro-Raman measurements of the bare Ge substrate, the Ge substrate with patterned Pd electrodes, the same sample in electrolyte, and after CV cycling confirm the stability of the Ge layer throughout these steps, see Fig.\ref{Figure S2} in Appendix A for details. 

Hydrogen absorption in the palladium film was carried out using the potentiostatic technique with chronoamperometry (CA) curves shown in Fig.\ref{Figure 2}(c), where a constant reduction potential (\qty{-0.26}{\volt} vs Ag/AgCl) was applied for hydrogen loading, followed by an oxidation potential (\qty{0.24}{\volt} vs Ag/AgCl) for desorption. The volume expansion scales with the atomic ratio of H/Pd [H/Pd]. The number of absorbed hydrogen atoms was derived from the reduction/oxidation charge by integrating the chronoamperometry absorption/desorption curve with background current removed, and the number of palladium atoms was calculated from the geometric volume (electrode area $\times$ film thickness) and density of the Pd film.

\begin{figure}[htbp]
    \centering
    \includegraphics[width=1\linewidth]{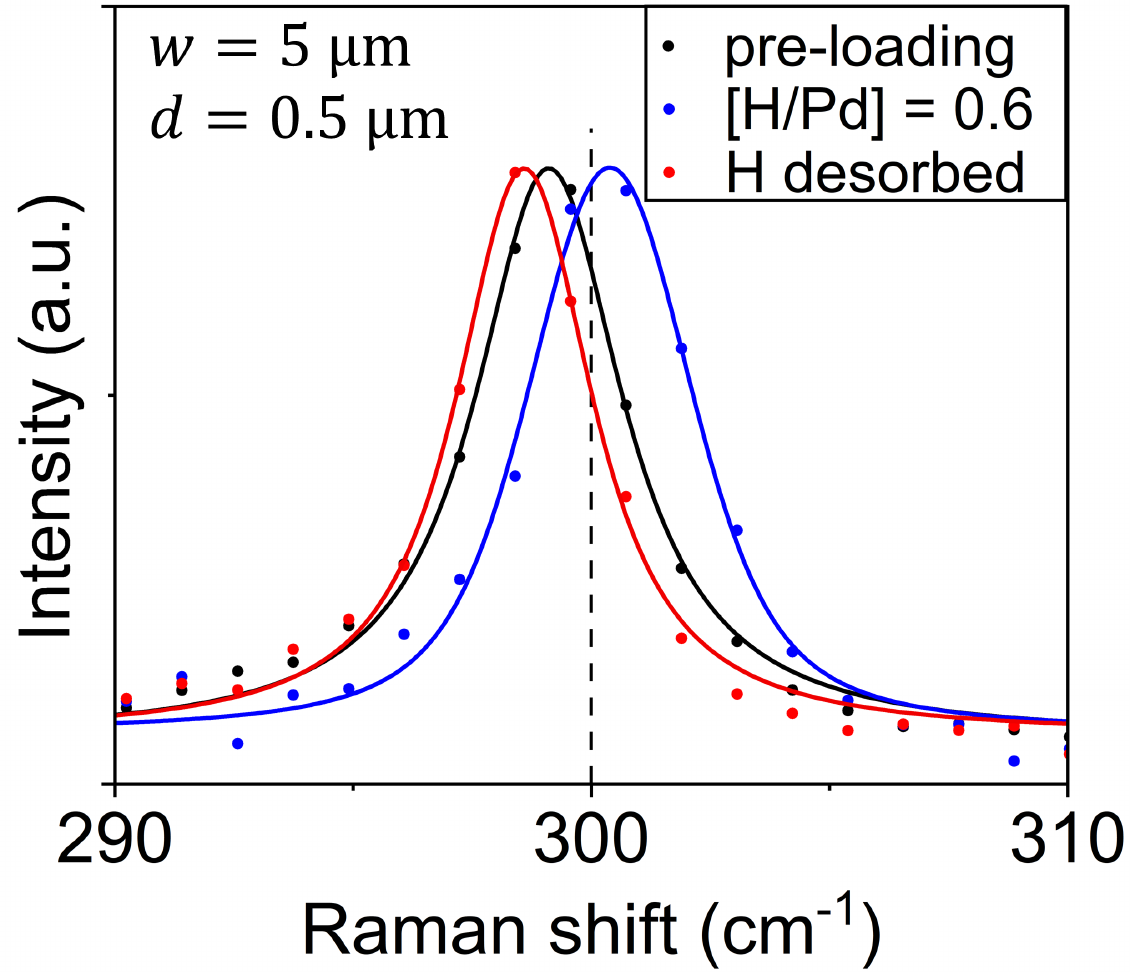}
    \caption{Raman measurement of stress induced in a Ge-on-Si film due to hydrostatic expansion of a structured Pd film due to H absorption. The black curve shows the measured Raman shift of the bare film with Pd electrodes pre-loading, the blue curve shows the shift at with a [H/Pd] loading ratio of 0.6 and the red curve shows the shift once the H is desorbed. All measurements are made at the same location at the centre of the electrode with parameters $w=$ \qty{5}{\um}, $d=$ \qty{500}{\nm}, cf. Fig.\ref{Figure 2}(a). The compressive stress induced by the expansion is clearly seen. The Raman shift of unstrained Ge is shown by the dashed line.} 
    \label{Figure 3}
\end{figure}

We use a micro-Raman spectrometer (Renishaw InVia) to in situ quantify the local strain induced by the volume expansion of the Pd electrode on the Ge film. The laser excitation wavelength in the system is \qty{488}{\nm}, and it is equipped with a 3-axis microscope stage enabling a mapping step size of \qty{100}{\nm}. The laser beam is focused on the sample through an objective lens with a numerical aperture (NA) of 0.6 and a 50$\times$ magnification, resulting in a spot size of $\approx$ \qty{700}{\nm} on the sample surface, while being focused through an electrolyte. Given the absorption length in Ge at the laser wavelength is $\approx$ \qty{20}{\nm} , the Raman (strain) signals reported here should be interpreted as the effective strain being induced in the top \qty{10}{\nm} of the Ge film.

From the measured Raman peak shift $\nu$, the strain induced in Ge relative to the unstrained Ge film can be derived from the Raman frequency shift $\Delta \nu=\nu-\nu_{0}=b\frac{(\varepsilon_{xx}+\varepsilon_{yy})}{2}$, with $b=[q-p(c_{12}/c_{11})]/{\nu}_0$. Here, $\nu_{0}$ is the frequency shift for bulk Ge (\qty{300}{\cm^{-1}}\cite{yoo2014characterization}), $p$ and $q$ are the phonon deformation potentials, $c_{11}$ and $c_{12}$ are the elastic constants of Ge. A value of $b=$ \qty{-415\pm 40}{\cm^{-1}} from \cite{fang2007perfectly} was employed in this study. To calibrate our measurements, we find a 0.2\% biaxial tensile strain using this approach on the unpatterned Ge film, which is consistent with the reported film strain due to thermal expansion mismatch between Ge and Si \cite{ishikawa2005strain, jain2011tensile}. Since the strain within the electrode gap induced by the Pd expansion is anisotropic, being predominantly compressive in the $y-$direction (cf. Fig.\ref{Figure 2}(a)), with a minor tensile component in the $x-$direction (shown in Fig.\ref{Figure S4} (c) and (d) in Appendix B), the strain is herein defined as the sum of the in-plane strain components $(\varepsilon_{xx}+\varepsilon_{yy})$. The strain change $\Delta(\varepsilon_{xx}+\varepsilon_{yy})$ in the following discussions refers to the net in-plane strain change relative to the Ge film with Pd electrodes before any electrochemical experiments are performed. We also assume that the $x,y$ axes in Fig.\ref{Figure 2}(a) correspond to the [100] crystal axis of Ge.

Fig.\ref{Figure 3} shows a representative Raman shift measurement to map the strain at the centre of the electrode gap, as shown in Fig.\ref{Figure 2}(a), with parameters $w=$ \qty{5}{\um} and gap $d=$ \qty{500}{\nm}. The black curve shows the measured shift of the device pre hydrogen loading, the blue curve is the shift with [H/Pd] = 0.6, and the red curve is taken after the H is desorbed from the device. The hydrostatic expansion induced Raman shift is clearly visible. As expected, and confirmed by finite element method (FEM) simulations below (Fig.\ref{Figure S4} in Appendix B), we observe a compressive strain in the region between the electrodes due to hydrostatic expansion in the electrodes themselves. To reiterate, the Ge film underneath the electrodes, which we cannot access in this measurement, is tensile strained (Fig.\ref{Figure S5}) but the inter-electrode gap (cf. Fig.\ref{Figure 2}(a)) is compressively strained.  

\begin{figure}[htbp]
    \centering
    \includegraphics[width=1\linewidth]{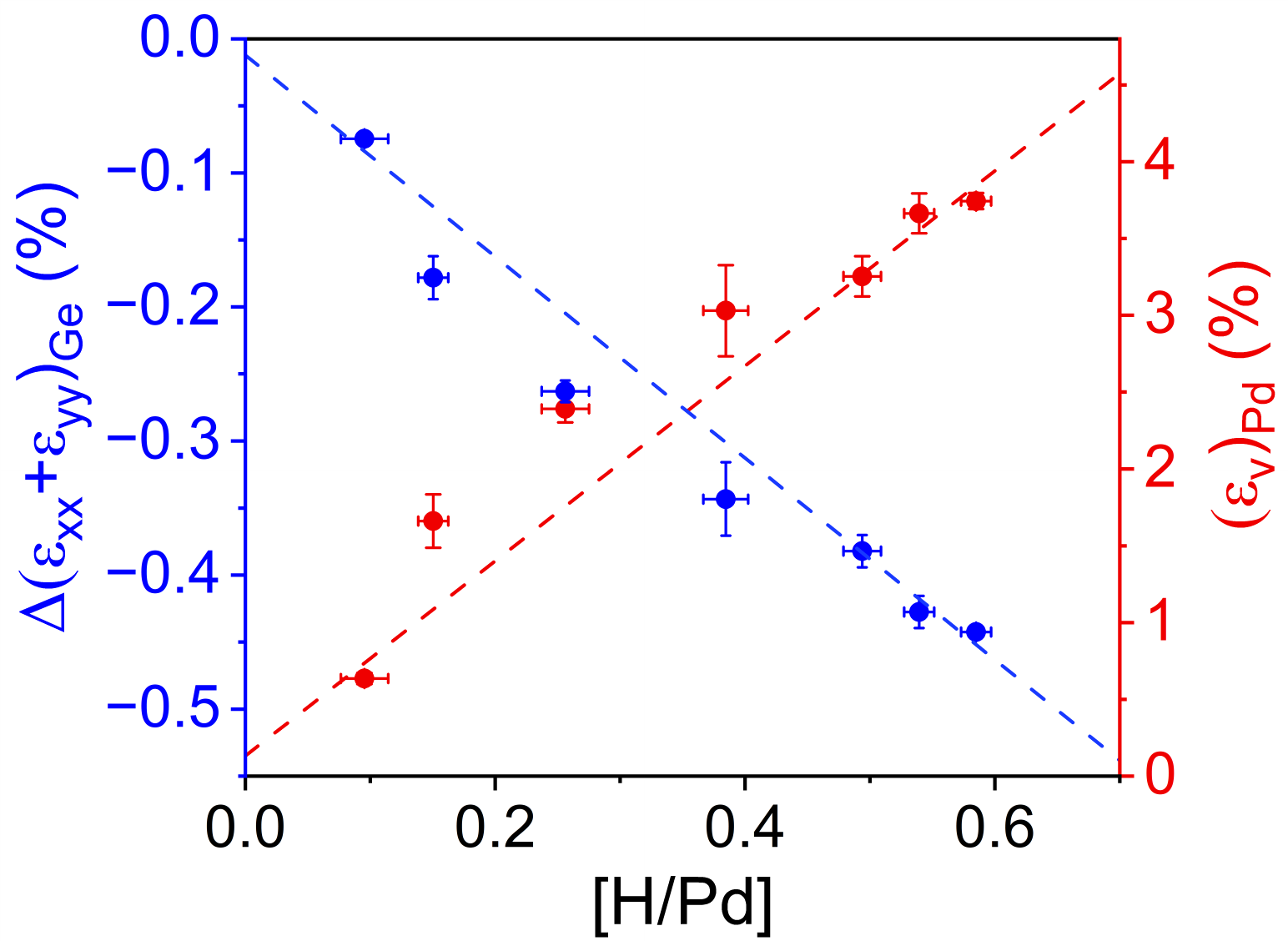}
    \caption{The left y-axis shows the induced in-plane strain change as a function of the H-loading fraction. The strain is extracted from the Raman shift as shown in Fig.\ref{Figure 3} for the Ge film at the centre of the electrode gap as indicated in Fig.\ref{Figure 2}(a) for a device with parameters $w=$ \qty{5}{\um} and gap $d=$ \qty{1}{\um}. The compressive strain increases approximately linearly (blue dashed line) with the loading fraction as expected. Using FEM modelling, one can use the measured strain to infer the volumetric expansion of the Pd electrodes which is shown with the right y-axis.} 
    \label{Figure 4}
\end{figure}

Fig.\ref{Figure 3} also shows that H loading and unloading process irreversibly transforms the Ge surface, indicated by the red curve (post H-desorption) not lining up with the black one (pre-loading H). We don't currently understand how the Ge surface is exactly modified after the repeated stress and de-stress cycles, but as we show below in Appendix C, the loading and unloading cycles do cause defect formation analogous to what has been observed in thermal cycling of Ge films between room temperature and \qty{850}{\celsius} \cite{persichetti2018formation}. We also note that the considerable lattice strain induced in the Pd film due to absorption of H in Pd near the $\alpha\to \beta$ phase transition of PdH\textsubscript{x} has been shown to result in defects and limited reversibility of Pd film-based actuators, which could potentially be playing a role here as well \cite{schoofs2009pd, verma2022dislocations}. 

As noted in the introduction, since the stress is controlled by the volumetric expansion of the Pd electrodes, which is determined by the [H/Pd] ratio, one can in principle program the stress precisely by controlling the [H/Pd] ratio. Fig.\ref{Figure 4} shows the measured in-plane strain change $\Delta(\epsilon_{xx}+\epsilon_{yy})$, which can be inferred from the Raman shift, as a function of the [H/Pd] loading fraction. The different data points in Fig.\ref{Figure 4}, corresponding to different [H/Pd] loading fractions, were taken over time during the potentiostatic hydrogen absorption. Each data point is an averaged value of five repeated Raman measurements taken at the centre of the electrode gap with parameters of $w=$ \qty{5}{\um} and $d=$ \qty{1}{\um}. The uncertainty in the [H/Pd] ratio arises from both the acquisition time ($\approx$ \qty{10}{\second} for the Raman shift overall), and uncertainty in the overall size of the electrode over which the H-loading occurs. The uncertainty due to acquisition time was estimated based on the [H/Pd] ratio change within the \qty{10}{\second} Raman measurement time interval from the chronoamperometry curve. For example, if the measurement starts at \qty{20}{\second}, [H/Pd] ratios at \qty{20}{\second} and \qty{30}{\second} can be derived from the chronoamperometry curve, and the difference between the ratios gives the uncertainty of each data point. 

\section{Modeling strain and inducing strain enhancement via geometry}

We use a finite element method modeling software, COMSOL Multiphysics,  to calculate the induced strain due to hydrostatic volume expansion of the Pd electrodes and set the model parameters according to the experimental conditions. The \qty{1.6}{\um} thick Ge layer and the Si substrate were defined as anisotropic elastic materials, while the \qty{200}{\nm} Pd electrode and \qty{2.5}{\nm} Cr adhesion layer were defined as isotropic. The dimensions of model were \qty{20}{\um} in width ($x$-axis in Fig.\ref{Figure 2}(a)) and \qty{60}{\um} in length ($y$-axis in Fig.\ref{Figure 2}(a)), with a fixed constraint applied to the bottom surface of Si substrate and all other surfaces free to deform. We add a background strain change of -0.1\% to the simulation at [H/Pd] loading of 0.6 as we observe the residual strain in our Raman measurements. In Fig.\ref{Figure 2}(a), we measure this residual strain at a location far (\qty{10}{\um}) from the inter-electrode gap. In theory, the (excess) strain there should be nearly zero, but is finite because of the effect of the unpatterned Pd electrode which surrounds the structured Pd region like a frame and is needed for maintaining electrical connectivity to drive the electroabsorption process.     \newline

\begin{figure}[htbp]
    \centering
    \includegraphics[width=1\linewidth]{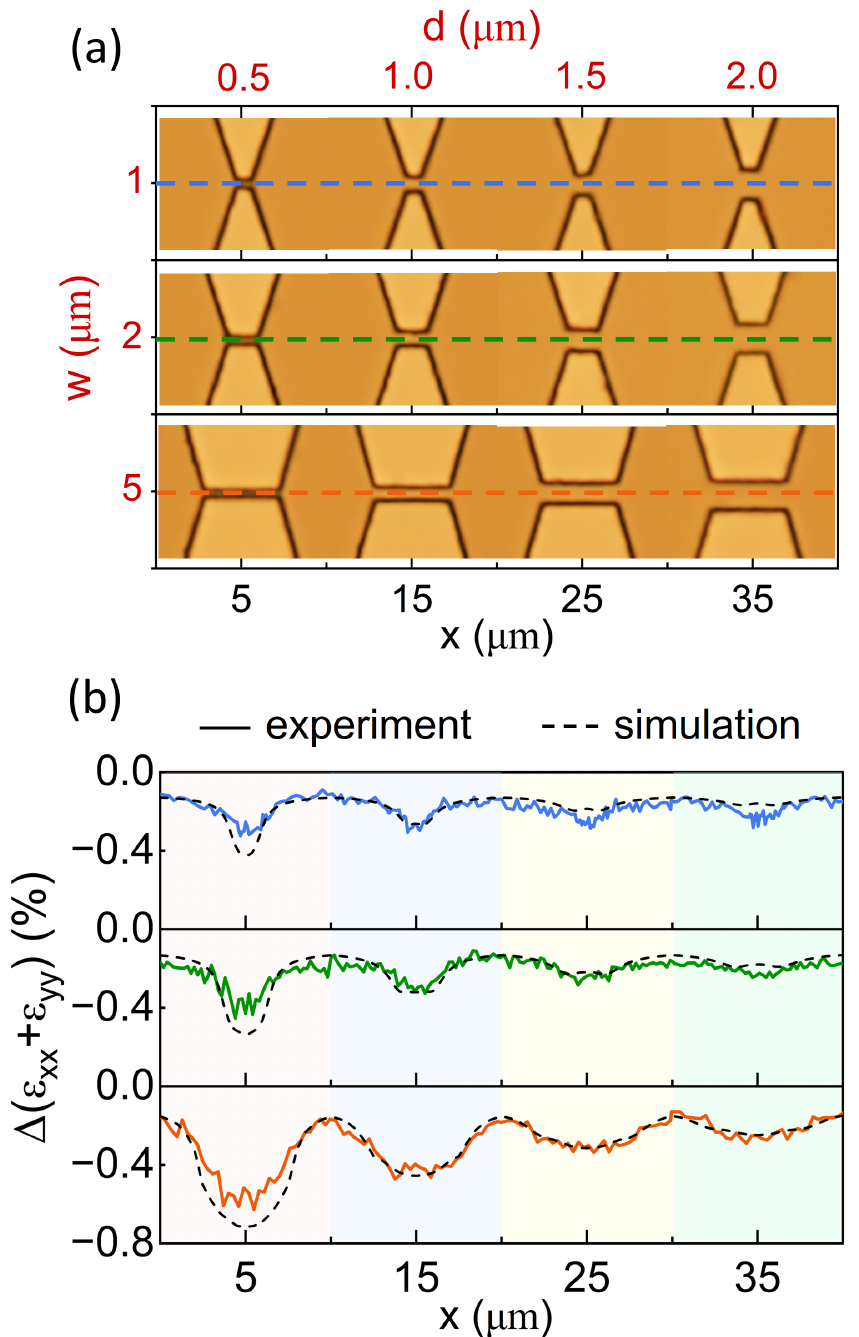}
    \caption{Controlling the 2D strain profile through varying electrode shape for a fixed [H/Pd] loading ratio of 0.6 (a) Optical micrograph of fabricated devices with fixed electrode widths $w$ of \qty{1}{\um} (top), \qty{2}{\um} (middle) and \qty{5}{\um} (bottom row) and varying electrode gaps $d$ ranging from \qty{0.5}{\um} to \qty{2}{\um}. (b) 1D strain maps (generated from the corresponding Raman data) of the devices from (a) taken across the centre of the inter-electrode gap. The effect of the electrode geometry on the induced strain is clearly seen and can be reproduced well from FEM simulations (dashed black curves). The discrepancy for the $d$=\qty{0.5}{\um} cases is discussed in the main text.} 
    \label{Figure 5}
\end{figure}

To get the volumetric strain $\varepsilon_{v, Pd}$, defined as the relative volume change ($\Delta V/V_{0}$) in the Pd electrode (right y-axis in Fig.\ref{Figure 4}), we iteratively adjust the strain in the Pd layer until we match the volume-averaged (laser spot size $\times$ absorption depth) strain change in the Ge film at the centre of the electrode. For instance, for electrode parameters of $w=$ \qty{5}{\um} and $d=$ \qty{1}{\um},  with a background strain ($\epsilon_{bkgd}$) of -0.1\% for a [H/Pd] loading ratio of 0.6, we find an $\varepsilon_{v, Pd}$ of 3.7\%. This values is about 3$\times$ lower than the 11\% lattice volume strain reported for both bulk Pd \cite{qian1988hysteresis} and a clamped Pd film \cite{pivak2011thermodynamics} for the similar [H/Pd] loading. However, it is important to keep in mind that for our structured films, the in-plane macroscopic expansion is more severely limited by the substrate constraint, resulting in an increased compressive stress and larger out-of-plane expansion in Pd film. Using our simulation mode, we estimate the the in-plane compressive stress in the Pd electrode, which we cannot access in the Raman experiment directly, and we find it to be $\approx$ \qty{-1.6}{\giga\pascal} at [H/Pd] = 0.6 (Fig.\ref{Figure S3}), which is in reasonable agreement with the reported value for a clamped film at similar [H/Pd] loading \cite{pivak2011thermodynamics}. Since we measure the residual background strain only at [H/Pd] = 0.6, the background corrections for the other [H/Pd] loading ratios in Fig.\ref{Figure 4} used to generate $\epsilon_{v,Pd}$, are linearly interpolated, so an $\epsilon_{bkgd}$ = -0.05\% is used for [H/Pd] = 0.3.

Fig.\ref{Figure 5} shows the effect of electrode shape on controlling the induced strain for a fixed [H/Pd] loading (constant $\epsilon_{v,Pd}$). It plots the 1D line-cut profiles of the measured and simulated strain along the x-axis at the centre of the inter-electrode gap. It plots the 1D line-cut $x$-profiles of the measured and simulated strain at the centre of the inter-electrode gap, shown by dashed lines in Fig.\ref{Figure 5}(a), for different electrode shapes (varying gap $d$, and width $w$) at a fixed [H/Pd] of 0.6. One can see that for a fixed gap $d$, the strain can be increased by increasing width $w$, and for a fixed $w$, the strain can be increased by reducing $d$. One can use this procedure to generate look-up tables for induced strain vs geometry for a given [H/Pd] ratio and then use this as a programmable knob to engineer the desired strain at a given location in the 2D-plane. The simulation methodology works well for predicting the induced strain at least for the cases where ($d,w>$ spot size). 

When the electrode gap $d$ becomes smaller than the spot size in our experiment (\qty{700}{\nm}), we find that the measured strain is significantly lower than what simulation predicts. This is especially clear in all the results for $d=$ \qty{0.5}{\um} in Fig.\ref{Figure 5}(b). We believe the discrepancy arises from change in the effective spot size on the Ge surface that is being sampled in the Raman measurement due to both the narrow gap (metal occlusion),  and the expansion of the Pd electrode. We tried to account for this in our simulation by reducing the effective spot size which was used to compute the volume averaged strain. The dashed black simulation curve for $d=$ \qty{0.5}{\um} in Fig.\ref{Figure 5}(b) was calculated using an effective spot size with width \qty{200}{\nm} along the $y-$axis and \qty{700}{\nm} along the $x-$ axis, cf. Fig.\ref{Figure 2}(a). Even with this reduction, there is a significant discrepancy where the measured strain is below what simulation predicts.  If we use a spot size of \qty{500}{\nm} along the $y-$axis corresponding to the actual electrode gap and the predicted laser spot size (\qty{700}{\nm}) along $x-$, the simulated compressive strain is $\approx$ -0.95\%, significantly above what we measure experimentally ($\approx$ -0.62\%).

\begin{figure}[htbp]
    \centering
    \includegraphics[width=1\linewidth]{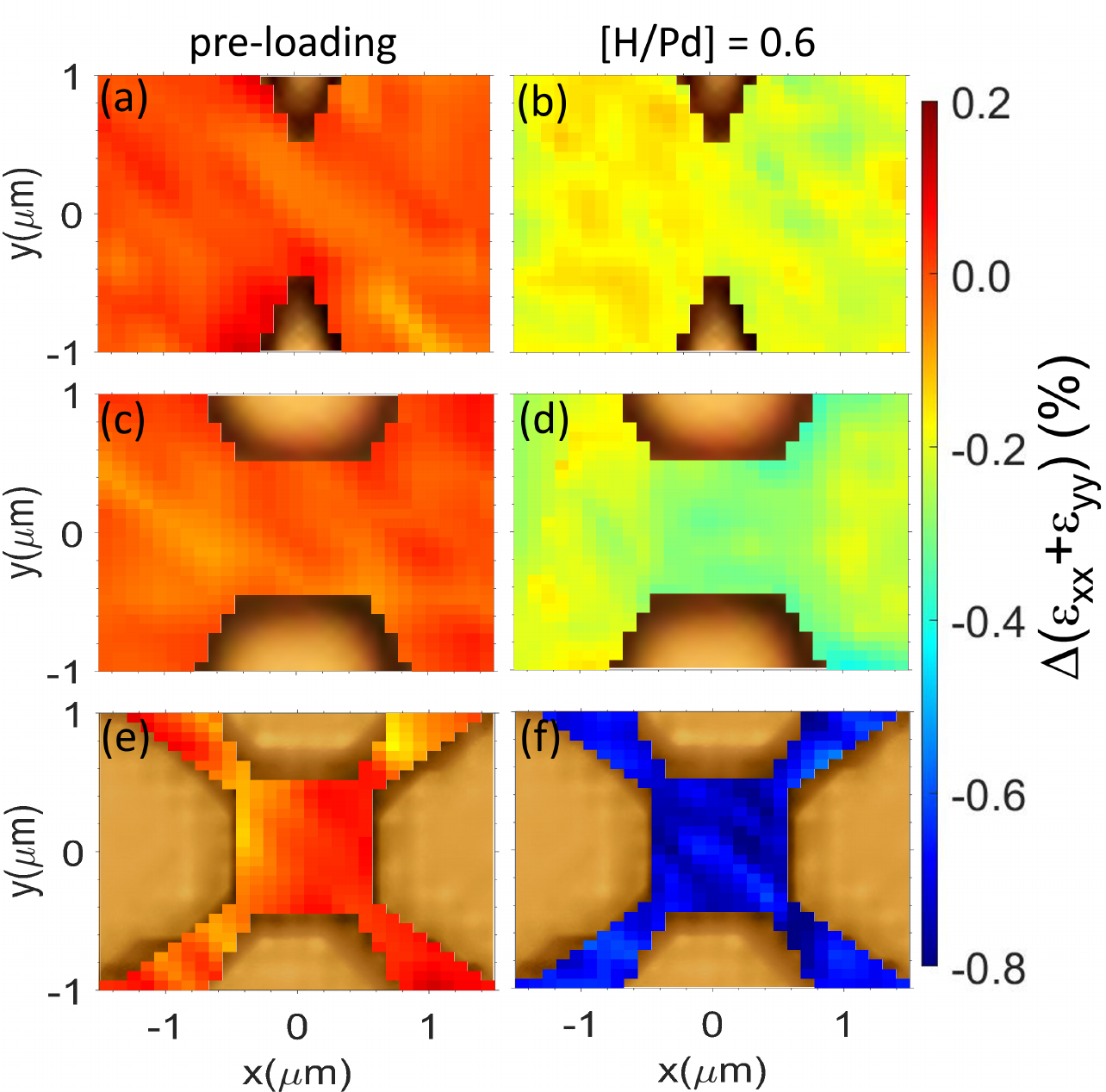}
    \caption{Strain concentration and enhancement via geometry. Since the Raman shift is sensitive to the total in-plane strain ($\epsilon_{xx}+\epsilon_{yy}$), one can shape the electrodes in order to exert a biaxial compressive strain which should double the overall measured strain. (a) electrode shaped to a point-tip with $w=0$, $d=$ \qty{1}{\um}, (b) standard electrode as discussed in Figs.\ref{Figure 3} and \ref{Figure 4} with $w=$ \qty{1}{\um} and $d=$ \qty{1}{\um}, and (c) a squared gap electrode geometry with electrodes along both the $x$ and $y$ axes with $w=$ \qty{1}{\um} and $d=$ \qty{1}{\um}. Doing a 2D strain scan (via the Raman shift) pre H-loading (left panel) and with a [H/Pd] = 0.6 (right panel) clearly shows the strain enhancement ($\approx$ 2$\times$) effect in the squared gap geometry in (c) compared to the reference case in (b).} 
    \label{Figure 6}
\end{figure}

The electrode geometry in Fig.\ref{Figure 5}(a) exerts primarily a compressive stress along the $y$-direction (cf. Fig.\ref{Figure 2}(a)) in the gap region. By adding a second pair of electrodes along an orthogonal direction (cf. Fig.\ref{Figure 6}), one can exert a biaxial compressive strain in the gap region and effectively measure double the induced strain due to the Raman measurement, which is sensitive to total in-plane strain ($\epsilon_{xx}+\epsilon_{yy}$). Assuming we limit the overall applied strain to operate within the linear regime, we can choose the in-plane electrode geometry to locally modify the stress and significantly enhance it by geometric stress concentration. An implementation of this ideas is shown in Fig.\ref{Figure 6} which shows the progressive increase in measured in-plane strain due to varying electrode geometry. The left side of Fig.\ref{Figure 6} (a,c,e) show the strain pre H-loading and the right side (b,d,f) show the strain for a loading [H/Pd] = 0.6 for different geometries. Moving from a point electrode ($w=$ 0) (b) to our standard reference electrode ($w=$ \qty{1}{\um}), the compressive strain in the gap increases, reproducing the results from Fig.\ref{Figure 5}. Adding a second electrode with ($w=$ \qty{1}{\um}) oriented orthogonally to the first (f), exerts a biaxial compressive strain due to expansion and the resulting Raman measured in-plane strain is significantly enhanced ($\approx$ 2$\times$) to -0.8\%. We would like to note that due to electrode symmetry, the strain is truly biaxial only at the exact centre of the gap.

\section{Improvements for set and forget programming of NOEM devices}
\begin{figure}[htbp]
    \centering
    \includegraphics[width=1\linewidth]{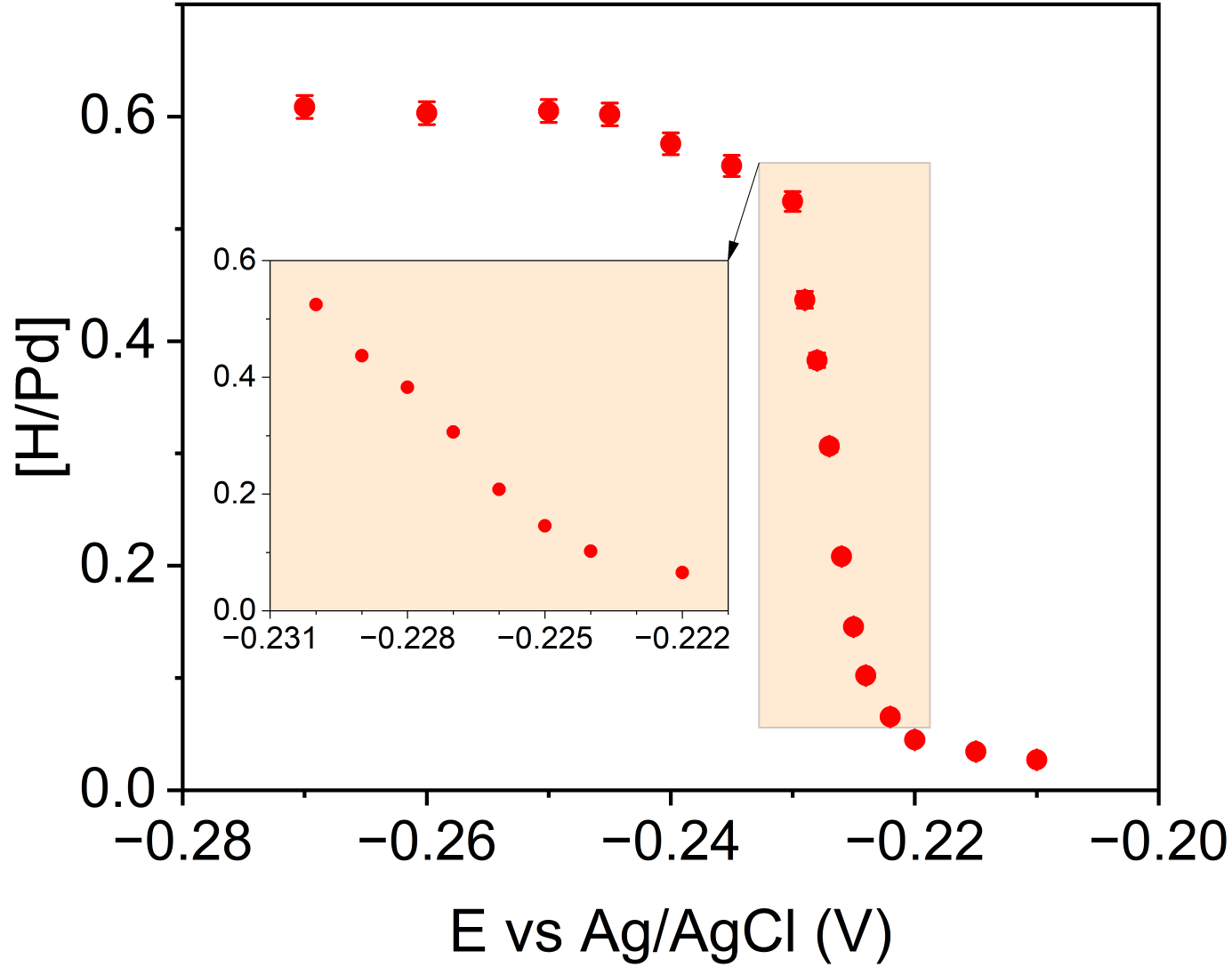}
    \caption{The dependence of [H/Pd] ratio, quantified through the oxidation charge from desorption CA curve, on the absorption potential. Each absorption potential was applied for at least \qty{10}{\min} until a stable background current obtained. Precision stress tuning using this technique requires precision control over the [H/Pd] loading ratio, which is limited in our experiment by the steepness of the loading curve with voltage, see inset.} 
    \label{Figure 7}
\end{figure}

While the proof-of-principle experiments demonstrated here convincingly demonstrate the potential of electrochemical approaches to generating controlled 2D strain patterns via the use of structured electrodes, there are a few challenges that need to be addressed if this technique is to be widely deployed for tuning NOEM devices at scale. While we have clearly shown that the combination of geometry (electrode shape) with the [H/Pd] loading ratio can be used as a platform for tunable strain generation, we haven't shown the precision control over local strain that the technique enables in principle. So, while we can tune over a variety of strains, we cannot set a fixed pre-determined value of strain at a location, as would be determined from a look-up table. We do the majority of the experiments in this work at [H/Pd] = 0.6 as the hydrogen in Pd film is saturated at 0.6. The tuning and fixation at lower [H/Pd] is challenging, as can be seen by the loading curve in Fig.\ref{Figure 7}, which shows the dependence of [H/Pd] ratio on the absorption potential. Each absorption potential was applied for a sufficiently long time, at least \qty{10}{\min}, until a stable background current was obtained. The saturated [H/Pd] ratios were quantified using the oxidation charges from desorption CA curves. The exquisite control over the [H/Pd] loading ratio, which is needed for precision stress engineering is possible in principle, but is limited in our experiment by the choice of \qty{0.1}{M} \ch{H2SO4} as the electrolyte and unavoidable presence of oxygen traces in the solution. The [H/Pd] loading ratio is extremely sensitive to potential as shown in Fig.\ref{Figure 7} due to the $\alpha\to \beta$ phase transition of PdH\textsubscript{x}. Moving to a different electrolyte, like an ionic liquid-based electrolyte \cite{hubkowska2024pd} will make this programming easier.

A second limitation on our experiments, especially with a view towards programming stress in NOEM devices, is the need to perform them in situ, as shown in the setup schematic in Appendix A. Currently, we can only maintain the H inside the Pd in an active electrochemical cell. As soon as the electrolytic reaction stops and the electrolyte is removed, the H desorbs from the Pd film and the applied stress is removed as the Pd relaxes back. For this technique to successfully interface with NOEM devices, the H needs to be trapped inside the Pd once a desired loading has been reached. This is critical for set-and-forget architectures where a desired strain can be programmed and maintained in a non-volatile fashion. Using sulphur to encapsulate the Pd film through surface poisoning has shown promise in this regard and is a key component of future studies \cite{castro2002effects,corthey2009electrochemical}.

\begin{figure}[htbp]
    \centering
    \includegraphics[width=1\linewidth]{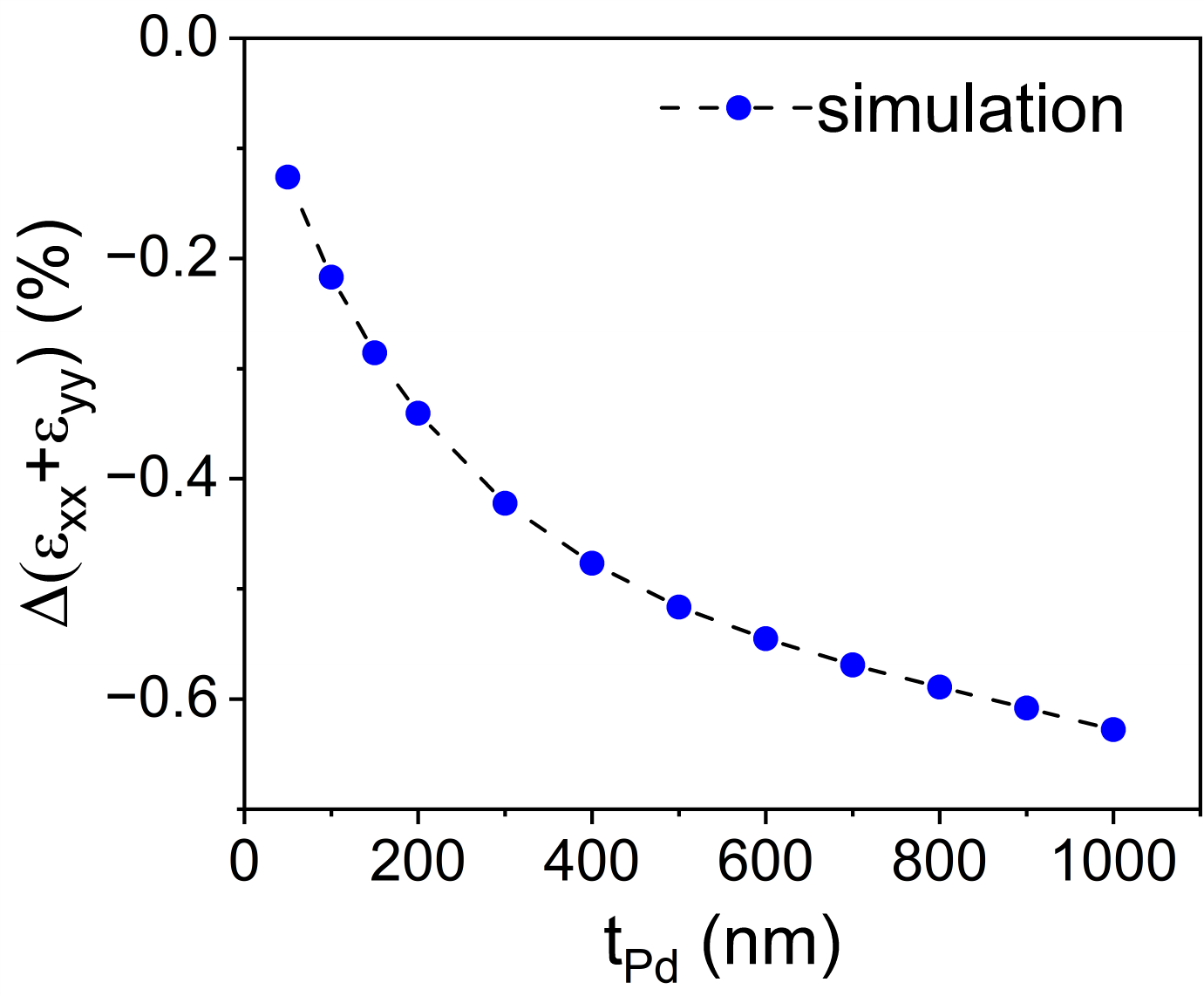}
    \caption{Dependence of the induced strain on the thickness of the Pd electrode for a fixed [H/Pd] = 0.6 estimated via FEM modelling. Simulations are carried out for the standard electrode geometry used in this work shown in Fig.\ref{Figure 2}(a) with electrode parameters $w$ = \qty{5}{\um} and $d$ = \qty{1}{\um}.} 
    \label{Figure 8}
\end{figure}

The other main open questions that need to be addressed moving forward, mainly with a view towards incorporating this tuning methodology with state-of-the-art NOEM devices is nanoscale electrode actuation and the limits on applied stress. As noted above, in our experiments while the electrodes are structured, we have a residual Pd metal frame around the array which ensures electrical connectivity. Ideally one wants to drive the electrochemical reaction on nanoscale Pd shaped electrodes so that the goal of 2D strain control can be realized. One way to circumvent the need for the Pd surrounding frame is to use the (doped) semiconductor as the electrode and study the effectiveness of inducing strain in the 2D plane using truly nanoscale electrodes, something that is currently infeasible with any other technique if one takes the full actuator footprint into account. 

Regarding the limits of strain that can be induced via this method, the two key parameters that limit the applied strain, ignoring any strain enhancements induced via geometric techniques as discussed above, are the [H/Pd] loading ratio and the thickness of the Pd film. While in this work, we saturate the H-loading at 0.6, Pd in principle can approach loading ratios > 0.8 \cite{benck2019producing} and a [H/Pd] ratio of 0.7 should be feasible in this thin-film platform. Increasing the Pd film thickness for a constant [H/Pd] ratio increases the induced strain. Fig.\ref{Figure 8} shows the increase in the in-plane strain for the standard electrode geometry, shown in Fig.\ref{Figure 2}(a), for device parameters $w$ = \qty{5}{\um} and $d$ = \qty{1}{\um}. Background strain was excluded here in line with the objective of nanoscale strain engineering discussed above. The net strain approaches -0.7\% for the single electrode case with Pd thickness of \qty{1}{\um} and with the enhancement effect shown in Fig.\ref{Figure 6} by incorporating a second set of orthogonally oriented electrodes, this can potentially give us in-plane compressive strains exceeding 1.2\%. In practice, we expect to keep the Pd film thickness below \qty{500}{\nm} in order to be compatible (reduce the overall mass-loading) with microelectromechanical systems (MEMS) based suspended devices wherein we envision such large strains to be applied without film damage. Moving towards suspended platforms is also critical for generating large volumetric strain transfers from the Pd to the underlying Ge film, as is needed for optoelectronic applications like bandgap engineering \cite{jain2011tensile, nam2011strained}. 

\section{Conclusions}

We have shown in this work that electrochemical techniques, combined with structured electrodes, can provide a viable route towards engineering arbitrary 2D strain patterns with the potential for engineering precision stress control in NOEM devices. We have shown some proof-of-principle experiments to validate the main ideas using the H-loading of Pd thin films and the resulting hydrostatic volume expansion to locally control the induce strain and show how the strain can be enhanced by a suitable choice of geometry. We have also outlined some of the main challenges that need to be addressed moving forward in order to make this technique feasible for \textit{set-and-forget} non-volatile strain programming in NOEM devices.

\section{Acknowledgements}

This work was supported in part by the UK’s Engineering and Physical Sciences Research Council (EP/X025381/1) and the UKRI frontier research guarantee (EP/Z000688/1). Nanofabrication was carried out using equipment funded by an EPSRC capital equipment grant (EP/N015126/1). DC acknowledges the support of the China Scholarship Council-University of Bristol PhD Scholarship (Award No. 202108270005). We thank David A.B. Miller for providing the Ge-on-Si substrates used in this work, and Laurent Kling, Martin Cryan, Robert Thomas, Alex Clark, Vivek Tiwari, Henkjan Gersen, Karen Grutter, Thomas Murphy, and John Rarity for valuable discussions and suggestions.

\section*{Appendix A: In situ Raman measurement: setup and data acquisition}

\begin{figure}[htbp]
    \centering
    \includegraphics[width=1\linewidth]{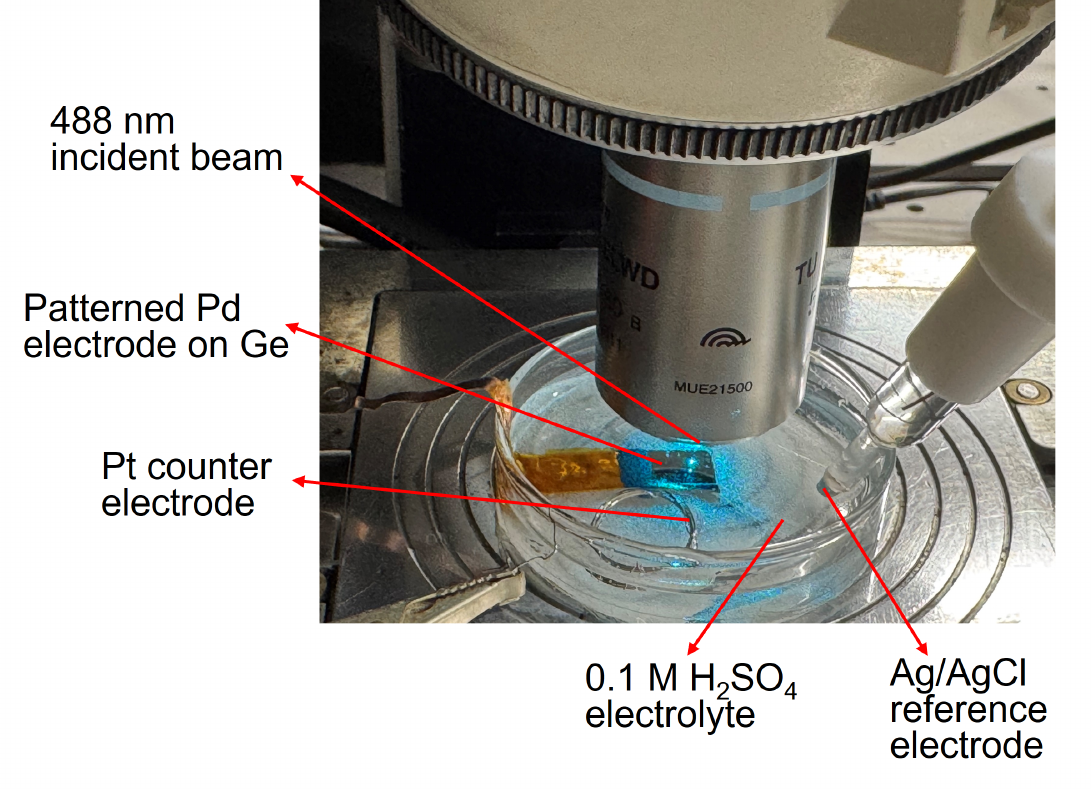}
    \caption{In situ Raman measurement setup: the sample is mounted in a home built electrolytic cell which can be mounted on the sample stage of the Raman spectrometer and fits within the working distance of the objective. The sample, electrodes and electrolyte are indicated.} 
    \label{Figure S1}
\end{figure}

\begin{figure}[htbp]
    \centering
    \includegraphics[width=1\linewidth]{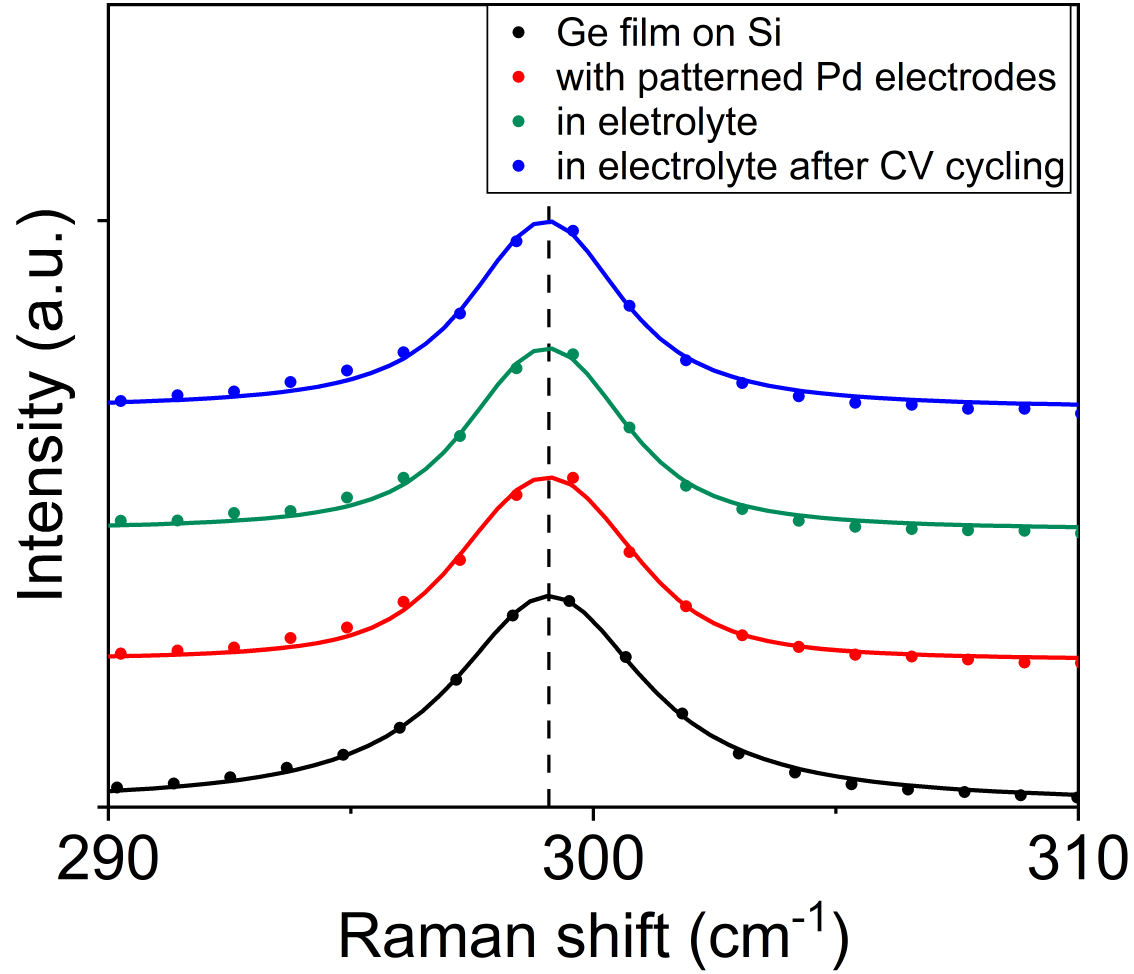}
    \caption{Raman measurements of the bare Ge-on-Si substrate, the Ge-on-Si substrate with patterned Pd electrodes, the sample in electrolyte, and after CV cycling. The last three measurements were taken on the exact same location.} 
    \label{Figure S2}
\end{figure}

The in situ Raman measurements were conducted in \qty{0.1}{M} \ch{H2SO4} using the setup shown in Fig.\ref{Figure S1} with Ag/AgCl as the reference electrode and a Pt wire as the counter electrode. The electrical contact on the sample was made using copper wire, sealed from the electrolyte with nail polish. The sealed area was minimized to maximize the accessible Pd film area, and the associated error was included in the geometric uncertainty for the [H/Pd] loading ratio calculation. During the Raman measurements, the laser power was adjusted to avoid heating on the sample and minimizing the effect on the measured current. A laser power of \qty{45}{\milli\watt} was applied: 1\% power with an acquisition time of \qty{8}{\second} for measurements in air, and 10\% with an acquisition time of \qty{2}{\second} in the electrolyte. Fig.\ref{Figure S2} shows the Raman shifts (three accumulations of each curve) of the bare Ge substrate, the Ge substrate with patterned Pd electrodes, the sample in electrolyte, and after CV cycling. The last three measurements were taken from the same sample at the same position, confirming the stability of the Ge layer throughout these steps. Although the fabrication-induced strain($\pm$ 0.1\%) in Ge film was observed in some samples, it was mostly localized at the edges of Pd electrodes. To minimize these edge effects, most of the measurements reported in this work with the exception of the 2D strain maps, were taken at or along the centre of the electrode gaps. 

The Raman data with hydrogen involved were collected after the first loading/unloading cycle as a non-linear strain-[H/Pd] relationship was observed during the initial loading, which we attribute to plastic deformation in the Pd electrode \cite{verma2022dislocations}, and the formation of defects in the Ge layer during the first loading. The strain response was consistent from the second loading onward, corresponding to that shown in Fig.\ref{Figure 4} in the main text, and no newly formed defects were observed in Ge layer as long as the [H/Pd] ratio was kept below 0.6. On the other hand, when we tried to increase the [H/Pd] ratio to $\approx$ 0.7 in \qty{0.5}{M} \ch{H2SO4}, we observed a drop in the  compressive strain during the fourth loading, accompanying the formation of new defects in the Ge layer. We discuss this further in Appendix C below. Figure \ref{Figure S9} shows the measured Raman shifts for intermediate [H/Pd] loading ratios. This data was used to generate the strain plot shown in Fig.\ref{Figure 4} in the main text.
  
\begin{figure}[htbp]
    \centering
    \includegraphics[width=1\linewidth]{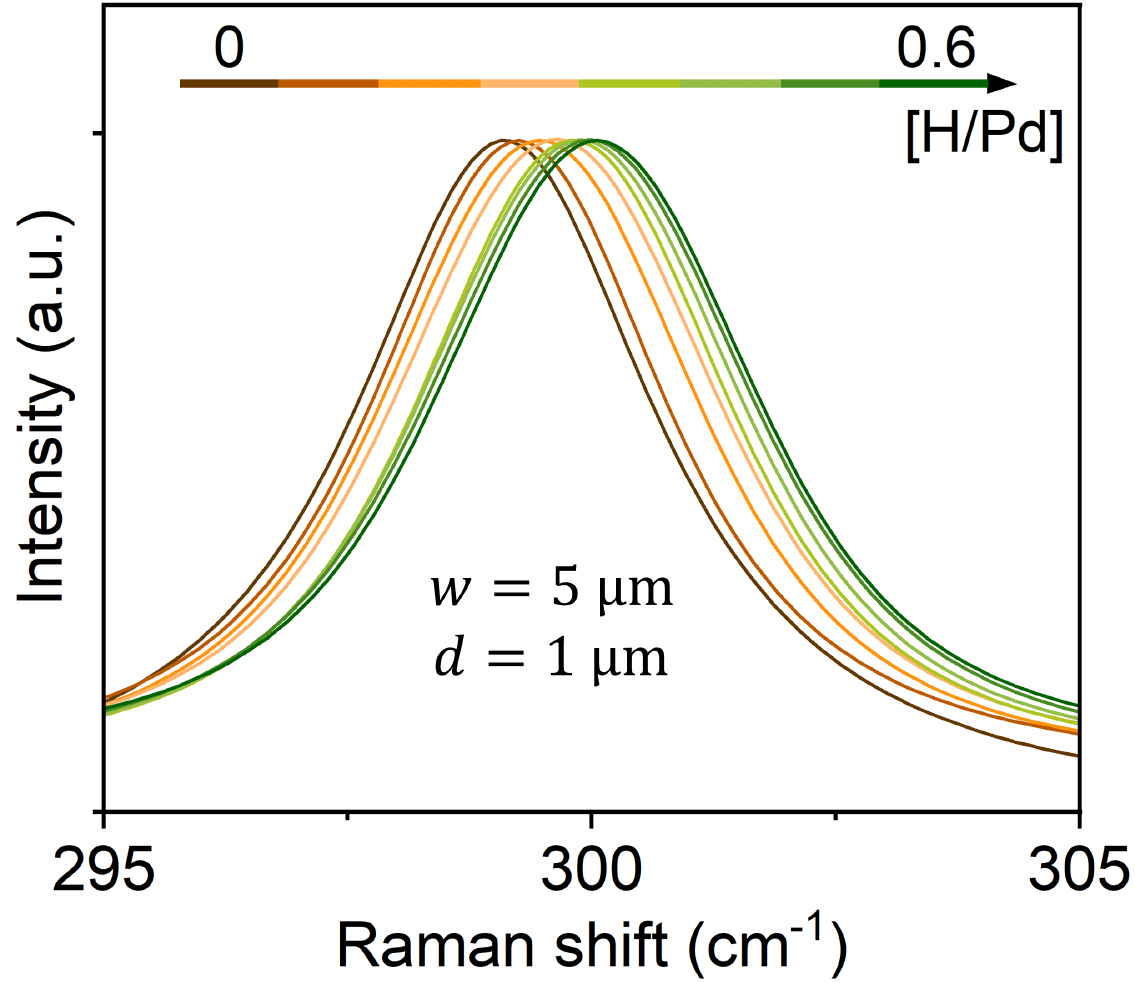}
    \caption{Measured Raman shifts for the standard electrode geometry $w$ = \qty{5}{\um} and $d$ = \qty{1}{\um} for intermediate [H/Pd] loading ratios. This data underlies the estimated strain values reported in Fig.\ref{Figure 4} in the main text.} 
    \label{Figure S9}
\end{figure}

\section*{Appendix B: Strain modelling using FEM simulations}

The FEM modelling, performed using COMSOL Multiphysics, consisted of two steps: first, we iteratively estimate the volumetric strain ($\varepsilon_{v, Pd}$) of the Pd electrode by reproducing the strain measured from the Ge film at the centre of electrode gap ($w$ = \qty{5}{\um}, $d$ = \qty{1}{\um}); then, we fix this extracted $\varepsilon_{v, Pd}$ and use it as an input (pre-stressed) to model the strain profiles that can be generated in other devices by varying electrode shapes. To compare the simulated strains with the strain extracted from the Raman measurements, an absorption volume of the laser (spot size $\times$ penetration depth) was estimated to calculate the average strain in the simulation. The \qty{488}{\nm} laser beam is focused on the sample through an objective lens with a numerical aperture (NA) of 0.6 and a 50$\times$ magnification, resulting in a spot size of $\approx$ \qty{700}{\nm} on the sample surface. This value is calculated from spot diameter $d=\frac{2\lambda}{\pi\cdot NA}$ with an assumed magnification factor of 1.4 for the measurements in electrolyte. The optical probing depth in Ge, relevant for the Raman measurements, is $\approx$ \qty{10}{\nm}.

\begin{figure}[htbp]
    \centering
    \includegraphics[width=1\linewidth]{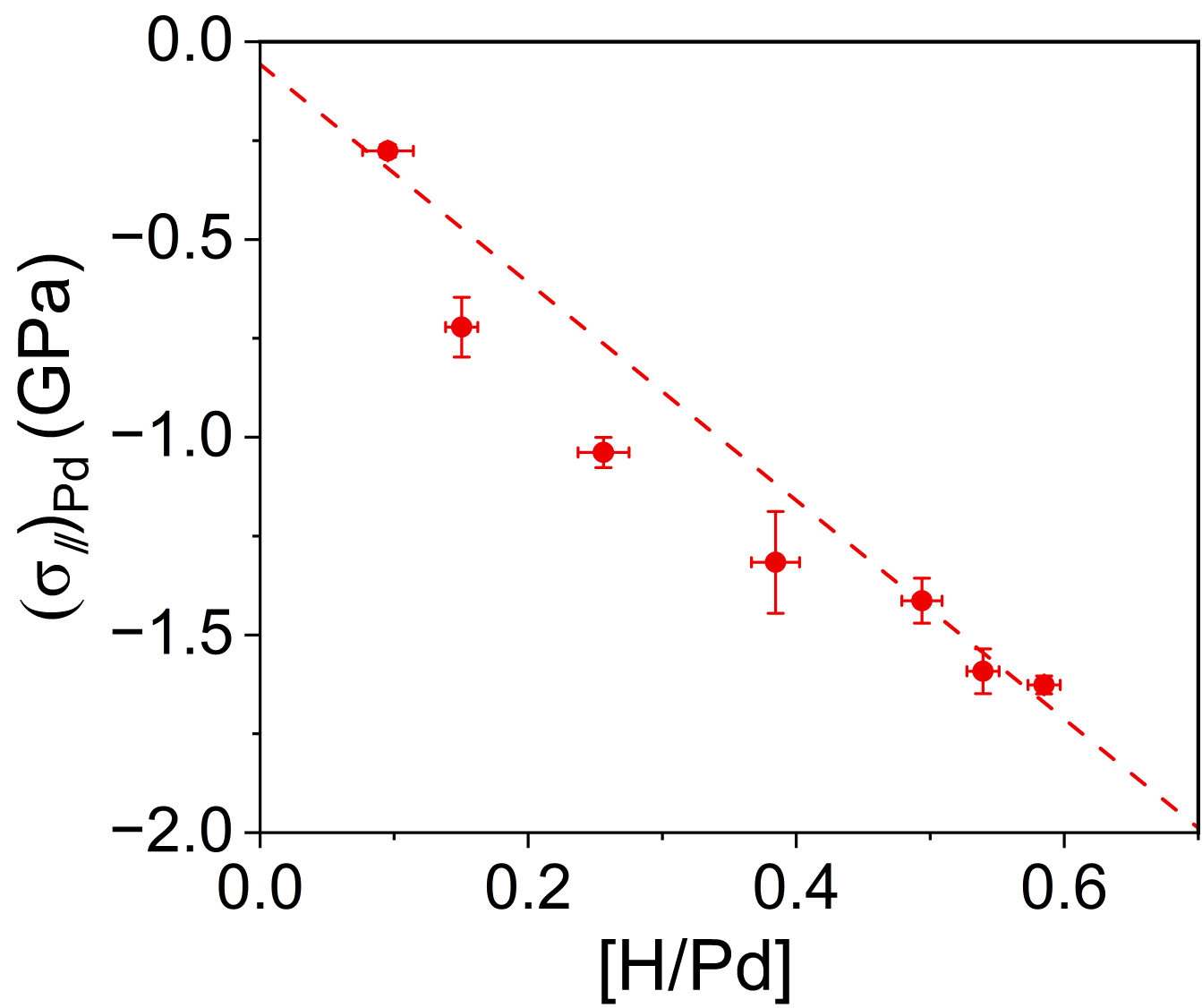}
    \caption{Predicted biaxial stress in Pd electrode ($w$ = \qty{5}{\um}, $d$ = \qty{1}{\um}) as a function of [H/Pd] ratio, iteratively derived from measured strain change $\Delta(\varepsilon_{xx}+\varepsilon_{yy})_{Ge}$ shown in Fig.4.} 
    \label{Figure S3}
\end{figure}

The \qty{1.6}{\um} thick Ge layer and the Si substrate were defined as anisotropic elastic materials, while the \qty{200}{\nm} Pd electrode and \qty{2.5}{\nm} Cr adhesion layer were defined as isotropic. The dimensions of model were \qty{20}{\um} in width ($x$-axis) and \qty{60}{\um} in length ($y$-axis), referred to the coordinate system in Fig.\ref{Figure 2}(a), with a fixed constraint applied to the bottom surface of the Si substrate and all other surfaces free to deform. The coordinate origin was defined as the centre of electrode gap on Ge surface and we assume that the coordinate axes are oriented along [100]. A background strain change of -0.1\% was applied to the simulations for [H/Pd] = 0.6 to get better agreement between simulations and experiment as we measure residual strain at a location of (\qty{10}{\um}) far from the inter-electrode gap. The strain there in theory (and predicted by the FEM simulation) is nearly zero and this background strain comes from the effect of the unpatterned Pd electrode which surrounds the structured Pd region like a frame and is needed for maintaining electrical connectivity to drive the electroabsorption process. Because of computational restrictions, we do not model this frame and instead add a background strain to our simulations. As noted in the main text, the background strain for other [H/Pd] loading ratios reported in Fig.\ref{Figure 4} was linearly extrapolated from the -0.1\% at [H/Pd] = 0.6. Fig.\ref{Figure S3} shows the derived biaxial stress in the Pd electrode ($w$ = \qty{5}{\um}, $d$ = \qty{1}{\um}).

\begin{figure}[htbp]
    \centering
    \includegraphics[width=1\linewidth]{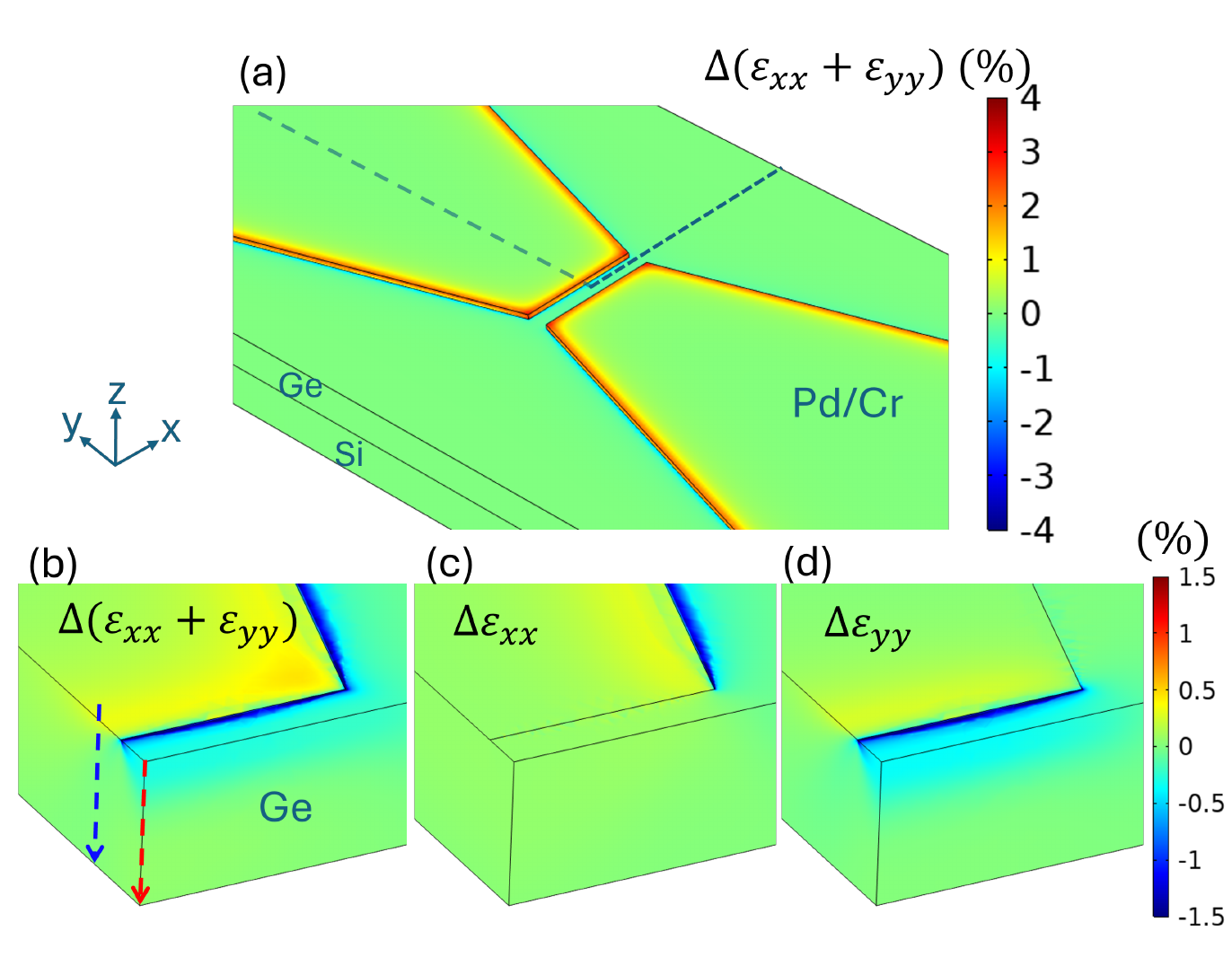}
    \caption{Simulated strain profiles with Pd electrode geometry of $w$ = \qty{5}{\um}, $d$ = \qty{1}{\um} with the origin of the coordinate system defined as the centre of electrode gap on the Ge surface. (a) $\Delta(\varepsilon_{xx}+\varepsilon_{yy})$ distribution for all layers. (b) $\Delta(\varepsilon_{xx}+\varepsilon_{yy})$, (c) $\Delta\varepsilon_{xx}$ and (d) $\Delta\varepsilon_{yy}$ profiles in the Ge layer sliced along $x=0$ and $y=0$, indicated by the dashed lines in (a).}
    \label{Figure S4}
\end{figure}

For the simulations to study the dependence of strain on Pd thickness ($t_{Pd}$) shown in Fig.\ref{Figure 8}, and strain profiles of device shown in Fig.\ref{Figure S4}, the background strain was excluded in line with our long-term objective of using nanoscale Pd electrode actuators strain engineering. Fig.\ref{Figure S4} shows the simulated strain profiles of a device with $w=$ \qty{5}{\um}, $d=$ \qty{1}{\um} with all layers. Fig.\ref{Figure S4}(b), (c), (d) show the zoomed-in strain profiles of $\Delta(\varepsilon_{xx}+\varepsilon_{yy})$, $\Delta\varepsilon_{xx}$ and $\Delta\varepsilon_{yy}$ of Ge layer sliced along $x=0$ and $y=0$, indicated by dashed lines in Fig.\ref{Figure S4}(a). At the centre point of the electrode gap, $\Delta(\varepsilon_{xx}+\varepsilon_{yy})=$ -0.30\%, $\Delta\varepsilon_{xx}=$ 0.06\% and $\Delta\varepsilon_{yy}=$ -0.36\% from the simulation, indicating the tensile strain in $x-$ direction and compressive strain in $y-$ direction. The volume averaged strains calculated from laser absorption volume are $\Delta(\varepsilon_{xx}+\varepsilon_{yy})=$ -0.34\%, $\Delta\varepsilon_{xx}=$ 0.06\% and $\Delta\varepsilon_{yy}=$ -0.40\%).

\begin{figure}[htbp]
    \centering
    \includegraphics[width=1\linewidth]{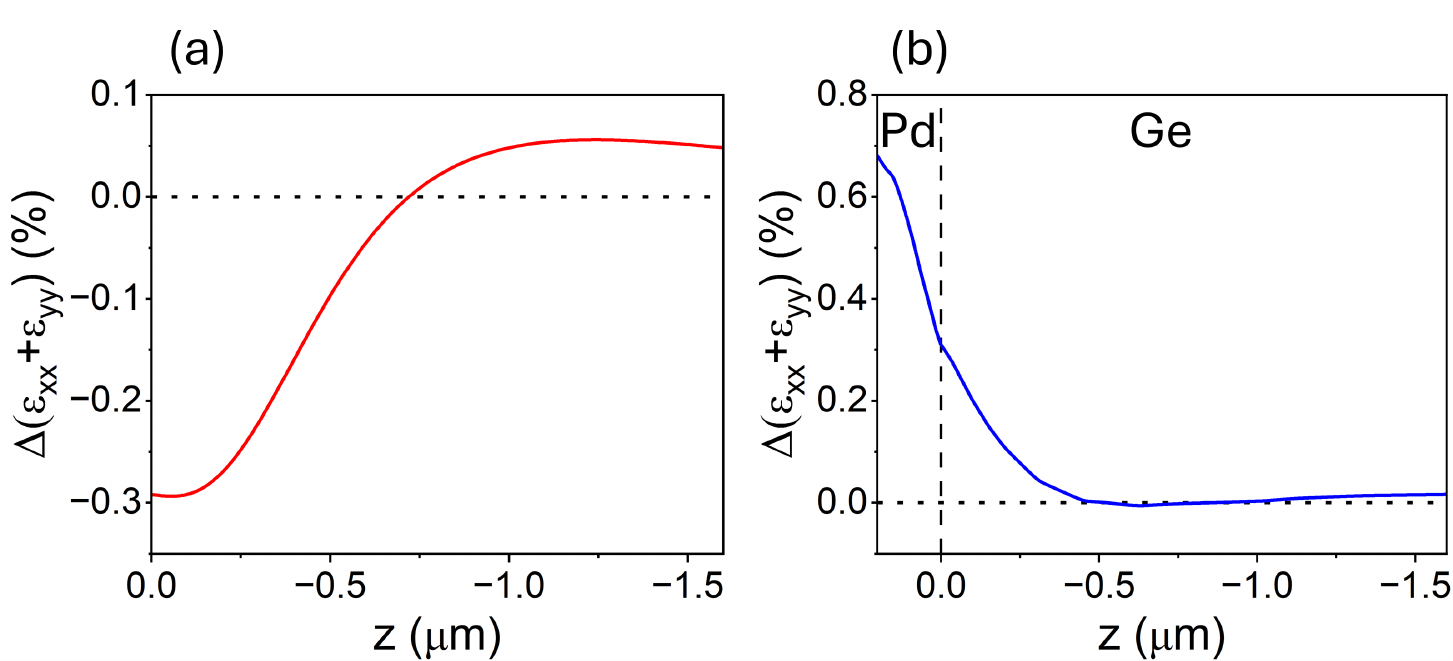}
    \caption{Evolution of strain with depth ($z$-profiles) shown by 1D strain cuts taken along the (a) red-dashed and (b) blue-dashed lines indicated in Fig.\ref{Figure S4}(b). $z=$ 0 is defined at the Ge surface. As noted in the main text, the in-plane strain under the electrodes is predominantly tensile, switching to compressive in the electrode-gap which is what we mainly measure in the Raman experiments. } 
    \label{Figure S5}
\end{figure}

Because of the penetration depth of the laser used in the Raman experiments, most of the strain values we report are effectively surface strains (depth $\approx$ \qty{10}{\nm}). To get an estimate of the strain evolution with depth, we rely on our FEM simulation. Fig.\ref{Figure S5} shows the extracted  $z-$profiles of strain (1D depth scan) along the dashed red and blue lines shown in Fig.\ref{Figure S4}(b). $z=$ 0 is defined at the Ge surface. Panel (a) corresponds to the red dashed line which is 1D strain depth scan taken at the centre of the electrode gap, and (b) the blue dashed line is measured \qty{0.5}{\um} away from the electrode edge under the Pd electrode. As expected, the in-plane strain switches polarity from being tensile under the electrodes to being compressive in the gap. In the measurements, we can only access the strain in the Ge film in the gap.

\section*{Appendix C: formation of defects}

\begin{figure}[htbp]
    \centering
    \includegraphics[width=1\linewidth]{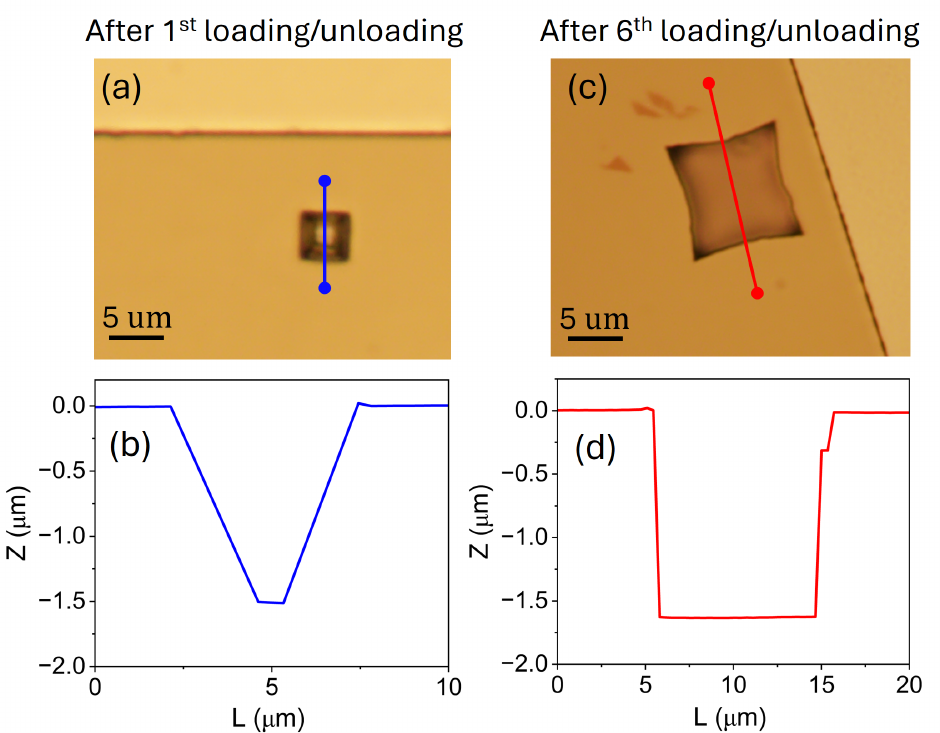}
    \caption{Optical microscope images of a representative defect formed in the Ge layer after the (a) first and (c) sixth loading/unloading cycle with thickness profiles along the blue (b) and red (d) line indicated underneath showing the Ge is effectively removed from that site.} 
    \label{Figure S6}
\end{figure}

\begin{figure}[htbp]
    \centering
    \includegraphics[width=0.8\linewidth]{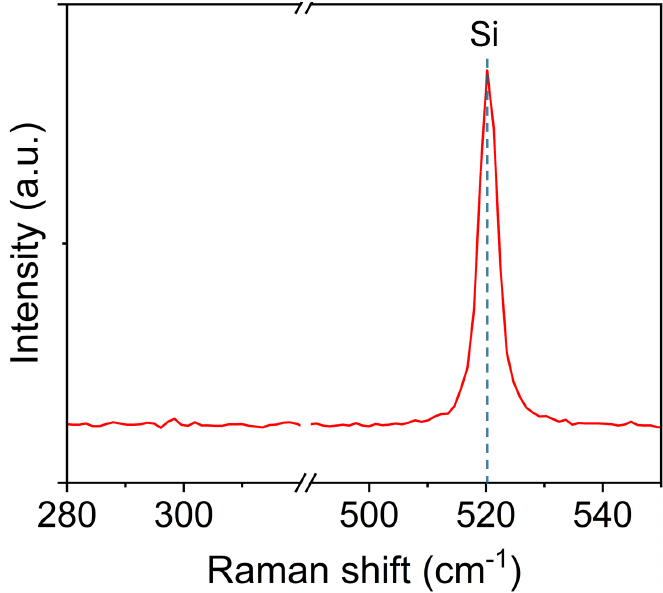}
    \caption{Measured Raman shift at the defect site shown in Fig.\ref{Figure S6} after the sixth loading/unloading cycle. We don't observe a Ge signal anymore further confirming that the Ge from that site is completely removed.} 
    \label{Figure S7}
\end{figure}

As noted in the main text, we observed defects forming in the Ge layer due to the repeated strain loading and unloading process. Fig.\ref{Figure S6} shows a representative microscope image of one of these defects formed in the Ge layer after the first (a) and sixth (c) loading/unloading cycles, giving a time-lapse image of defect evolution. Fig.\ref{Figure S6}(b) and (d) present the corresponding thickness profiles measured using an optical profilometer, showing defect depth of approximately \qty{1.6}{\um}, which matches the thickness of the Ge layer and shows that the defect there extends all the way to the Si substrate and the Ge is completely removed. This is further confirmed by the Raman measurement at the defect site from sample after the sixth loading/unloading cycle (Fig.\ref{Figure S7}), which shows only the Si signal. As noted in the main text, similar defects are observed in temperature cycling Ge-on-Si substrates between room temperature and \qty{850}{\celsius} \cite{persichetti2018formation}, which provides an indication of the stresses involved.

\begin{figure}[htbp]
    \centering
    \includegraphics[width=1\linewidth]{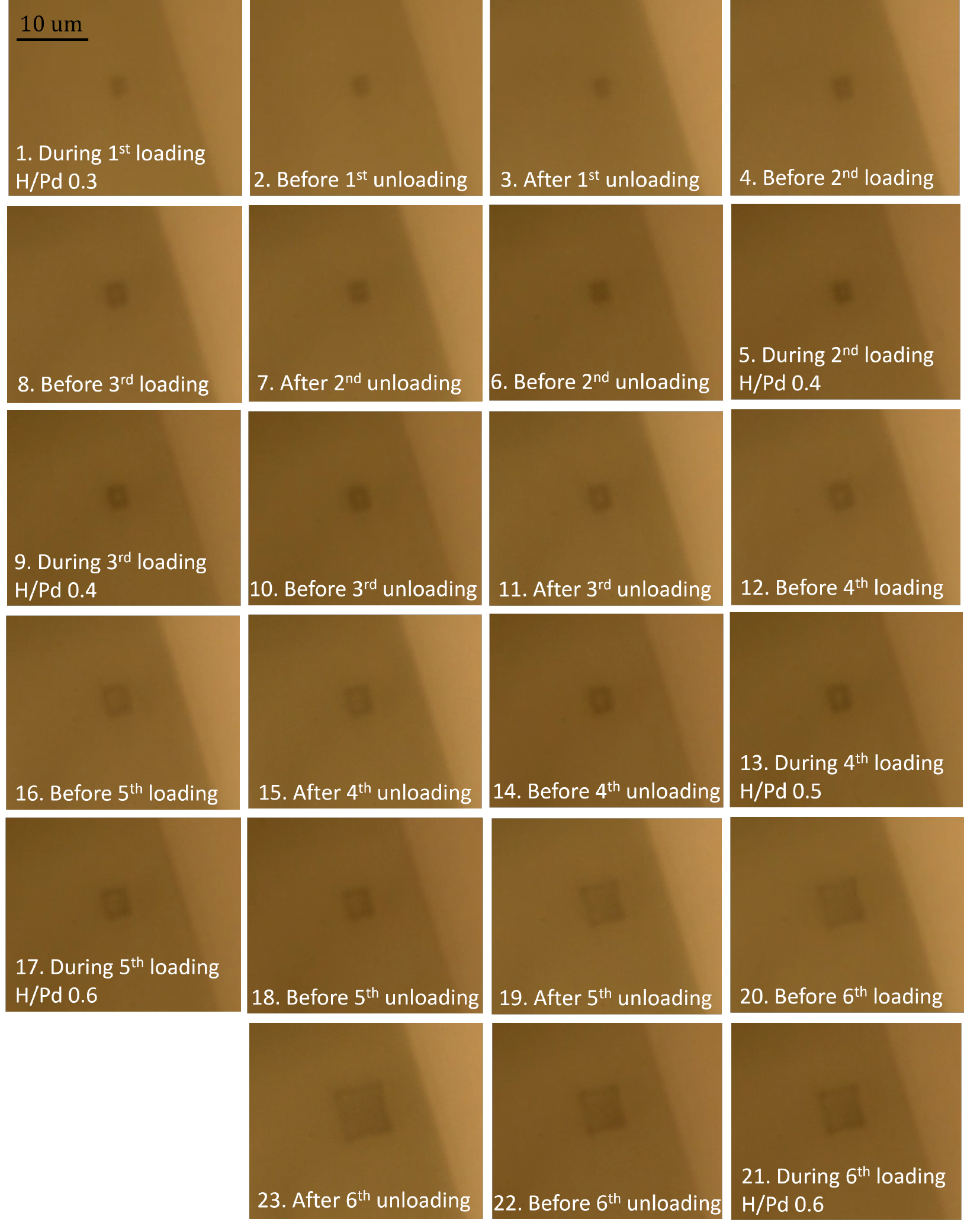}
    \caption{The evolution of a representative defect in the Ge layer as the sample is put through repeated H loading/unloading cycles. As the strain is cycled, the defect starts to grow in size. As can be seen, the defect is oriented along specific crystal axes pointing to the favorable energetics of defect formation for film relaxation under high stress. We empirically observe a higher concentration of defects near the electrode edges where the induced strain is significantly higher.} 
    \label{Figure S8}
\end{figure}

To better understand the enlargement of these strain induced fractures, we recorded the defect evolution during the loading/unloading cycles, as shown in Fig.\ref{Figure S8}. The fracture was formed during the first loading with [H/Pd] = 0.3. From the resolution of our microscope images, it is hard to pinpoint the exact [H/Pd] ratio at which it starts forming. The enlargement of the defect occurred primarily during the unloading, continuing even after the electrochemical unloading had stopped. We believe these defect likely originate from point defects / threading dislocations in the Ge film which become energetically favored under applied stress. The Ge around the dislocations was preferentially removed due to the strain concentration at the defect. Since the stain is more concentrated at the electrode edges as shown in Fig.\ref{Figure S4}(b), we find that these defects are more likely to form near the electrode edges, although we haven't statistically quantified this. We would like to note that the defects have a very specific crystalline orientation, as can be seen by their shapes, which also provides evidence for them becoming energetically favored under large applied stress, providing a mechanism for the film to relax. Moving to a membrane geometry provides additional routes for stress incorporation and potentially avoid these defects forming. In prior experiments with membranes \cite{jain2011tensile, nam2011strained} where similar strains were applied, we did not observe these defects although it is important to note that in those experiments the strain was fixed and not cycled unlike here.

%\clearpage
\bibliography{References}% Produces the bibliography via BibTeX.

\end{document}